\documentclass[prd,aps,preprintnumbers, showpacs, preprint, nofootinbib,superscriptaddress]{revtex4-1}
\usepackage{epsfig}
\usepackage{bm}
\usepackage{amssymb}
\usepackage{amsmath}
\usepackage{color}
\usepackage{subfigure}
\newcommand{\be}{\begin{equation}}
\newcommand{\ee}{\end{equation}}
\newcommand{\bea}{\begin{eqnarray}}
\newcommand{\eea}{\end{eqnarray}}

\newcommand{\beq}{\begin{eqnarray}}
\newcommand{\eeq}{\end{eqnarray}}
\newcommand{\nn}{\nonumber \\}

\begin{document}

\title{Flow harmonics $v_n$ at finite density}

\preprint{YITP-15-39}

\author{Yoshitaka Hatta}
\affiliation{Yukawa Institute for Theoretical Physics, Kyoto University, Kyoto 606-8502, Japan}

\author{Akihiko Monnai}
\affiliation{RIKEN BNL Research Center, Brookhaven National Laboratory, Upton 11973 NY, USA}

\author{Bo-Wen Xiao}
\affiliation{Key Laboratory of Quark and Lepton Physics (MOE) and Institute
of Particle Physics, Central China Normal University, Wuhan 430079, China}

\date{\today}

\begin{abstract}
 We investigate the Gubser solution of viscous hydrodynamics at finite density  and analytically compute the flow harmonics $v_n$. We explicitly show how $v_n$ and their viscous corrections depend on the chemical potential. The difference in  $v_n$ between particles and antiparticles is also analytically computed and shown to be proportional to various chemical potentials and the viscosity. Excellent agreement is obtained between the results and the available experimental data from the SPS, RHIC and the LHC.
\end{abstract}
\maketitle

\section{Introduction}

Relativistic hydrodynamics is a general theoretical framework to describe the collective dynamics of high-energy systems near local thermal equilibrium. Its first application to hadron physics dates back to Landau's attempt to describe multi-particle production in hadron-hadron collisions \cite{Landau}. It has become a topic of great interest since the discovery of the quark-gluon plasma (QGP) as a nearly-perfect fluid in the ``Little Bangs" at BNL Relativistic Heavy Ion Collider (RHIC) \cite{Adcox:2004mh, Adams:2005dq, Back:2004je, Arsene:2004fa} and CERN Large Hadron collider (LHC) \cite{Aamodt:2010pa, ATLAS:2011ah, Chatrchyan:2012wg}. This is supported by the observations that the azimuthal momentum anisotropy of hadronic distribution \cite{Ollitrault:1992bk, Poskanzer:1998yz}, characterized by flow harmonics $v_n$, are found to reflect the geometrical anisotropy  $\epsilon_n$ of the overlapping region of two colliding nuclei, and that they are in good quantitative agreement with theoretical estimations. Nowadays the viscous hydrodynamic modeling is considered as one of the most powerful tools to quantify the QGP medium near the crossover phase transition \cite{Schenke:2010rr}.

The recent Beam Energy Scan (BES) experiments at RHIC pose us intriguing challenges to study the properties of the medium at finite density and to explore the QCD phase diagram to find signs of a critical point \cite{star}. Conserved charges such as net baryon number, strangeness and isospin would play important roles in the collisions with lower energies, as the differences between particle and antiparticle yields are clearly seen \cite{Adamczyk:2013gv,Adamczyk:2013gw}. Historically, it had long been speculated based on several idealized calculations that the strong coupling limit is achieved only at highest energies of RHIC experiments. On the other hand, recent improvements in off-equilibrium hydrodynamic modeling motivates us to reexamine the validity of  hydrodynamics in exploring the dense quark matter created at mid-low energies, especially since the differential elliptic flow $v_2 (p_T)$ is found to remain large in Phase~I of the BES experiments. The applicability of hydrodynamic models is closely related to the origin of fluidity, about which little is known, and thus its verification would be a very important step towards a full understanding of the hot QCD medium.

So far many hydrodynamic analyses have been performed numerically because it is generally quite  nontrivial to solve the partial differential equations involved. Analytical solutions of relativistic hydrodynamics, on the other hand, can be obtained with certain symmetry conditions and they are very instructive in understanding the essence of heavy-ion dynamics. The boost-invariant Bjorken flow \cite{Bjorken:1982qr} is one such classic example. More recently,  Gubser found an exact boost-invariant solution of the Navier-Stokes equation which has a nontrivial dependence on the transverse coordinate \cite{Gubser:2010ze}. The latter solution has the advantage that one can add azimuthally anisotropic perturbations \cite{Gubser:2010ui,Hatta:2014upa} and analytically compute the corresponding flow harmonics $v_n$ including the viscosity effects \cite{Hatta:2014upa,Hatta:2014jva} (see, also, \cite{Csanad:2008af,Peschanski:2009tg,Csanad:2014dpa}).

In this study, we  investigate $v_n$ at finite density by analytically solving the viscous hydrodynamic equations coupled with conserved currents assuming  conformal and boost-invariant symmetries. Aside from the fact that the solution itself is new and of theoretical importance, it gives us a theoretical guidance about the behavior of $v_n$ over
a wide range of the beam energy for which there are not many numerical simulations \cite{Kolb:2000sd,Teaney:2001av,Petersen:2009vx,Shen:2012vn,Karpenko:2015xea} 
and the previous knowledge obtained through the precision analyses in the RHIC-LHC energy regime, such as the value of the shear viscosity $\eta/s$, are no longer fully applicable. We discuss extensively the nature of flow in the presence of currents and estimate the beam energy (or chemical potential) dependence of $v_n$. The difference in $v_n$ between particles and antiparticles is also analytically computed.
 The results are compared
with the experimental data from SPS, RHIC and the LHC \cite{Alt:2003ab,Adamczyk:2013gv,Adamczyk:2013gw,CMS:2013bza}. We see that they are in qualitative agreement, which suggests that a reasonable description of the low-energy experimental data might be possible within a hydrodynamic framework.

The paper is organized as follows. The basic setup of relativistic hydrodynamics is outlined in Section \ref{sec:2}. We then present analytical formulas of the flow harmonics $v_n$ in the ideal and viscous cases in Sections \ref{sec:3} and \ref{sec:4}, respectively. Phenomenological inputs for our model are summarized in Section \ref{sec:5}.
 Using these formulas and input parameters, we compare our results with the experimental data in Section \ref{sec:6}. Section VII is devoted to summary and conclusions.

\section{Hydrodynamic equations}
\label{sec:2}

\subsection{Setup}
We shall consider hydrodynamics of a conformal theory. The system is characterized by the local temperature $T$ and a set of local chemical potentials $\mu_i$ where the subscript $i$ labels various conserved charges of the theory. The flow velocity is denoted by $u^\mu$ with the normalization $u^\mu u_\mu=-1$.  The energy-momentum tensor in the Navier-Stokes approximation takes the form
 \beq
 T^{\mu\nu}=
 \frac{4\varepsilon}{3} u^\mu u^\nu + \frac{\varepsilon}{3}g^{\mu\nu} -2\eta \sigma^{\mu\nu}\,, \label{eosT}
 \eeq
  where $\sigma^{\mu\nu}$ is the shear tensor and $\eta$ is the shear viscosity. In (\ref{eosT}), the conformal equation of state $\varepsilon=3p$ between the energy density $\varepsilon$ and the pressure $p$ has been used.  The conserved current $J_i^\mu$ associated with the chemical potential $\mu_i$
 can be written as
\beq
J_i^\mu=n_iu^\mu -\kappa_i (u^\mu u^\nu+g^{\mu\nu})\partial_\nu \left(\frac{\mu_i}{T}\right)\,,
\label{charge}
\eeq
 where $n_i$ is the charge density and $\kappa_i$ is the charge conductivity.
The hydrodynamic equations consist of the conservation equations for $T^{\mu\nu}$ and $J^\mu_i$
\beq
\nabla_\mu T^{\mu\nu}&=&0\,, \qquad  \nabla_\mu J_i^\mu=0\,,
\label{current}
\eeq
 where $\nabla_\mu$ is the covariant derivative.

 Since there is no intrinsic mass scale in a conformal theory, the energy density $\varepsilon$ and the charge densities $n_i$ can be generically written as
\beq
\varepsilon=T^4f\left(\frac{\mu_1}{T},\frac{\mu_2}{T},\cdots\right)\,, \qquad
n_i=\mu_{i} T^2g_i\left(\frac{\mu_{1}}{T},\frac{\mu_2}{T},\cdots\right)\,.
\eeq
With a view to applying to heavy-ion collisions, we shall focus on the following representative situation.  We assume that there is the leading current $J^\mu=n u^\mu+\cdots$ (`baryon number current') and the corresponding chemical potential $\mu$  is treated to all orders. In addition, there is one subleading current $\tilde{J}^\mu=\tilde{n}u^\mu+\cdots$ (`isospin number current') whose chemical potential $\tilde{\mu}$ is small and treated only to linear order. We take $\tilde{\mu}$ to be `orthogonal' to $\mu$, in that  $\varepsilon(\mu,\tilde{\mu})$ is invariant under a sign flip $\tilde{\mu}\leftrightarrow -\tilde{\mu}$ (i.e., cross terms like $\mu\tilde{\mu}T^2$ are absent). With these assumptions, we can parameterize
\beq
 \varepsilon= T^4f\left(\frac{\mu}{T}\right)\,, \qquad
n=\mu T^2g\left(\frac{\mu}{T}\right)\,, \qquad
\tilde{n}=\tilde{\mu} T^2 \tilde{g}\left(\frac{\mu}{T}\right)\,. \label{eos}
\eeq
  The last equation may be written as $\tilde{n}=\tilde{\mu}\tilde{\chi}$ where  $\tilde{\chi}\propto \left.\partial^2 p/\partial \tilde{\mu}^2\right|_{\tilde{\mu}=0}$ is the susceptibility.

\subsection{Gubser flow}

We shall solve the hydrodynamic equations (\ref{current}) for a given flow  velocity
\beq
u^\tau=\cosh \left[\tanh^{-1}\frac{2\tau x_\perp}{L^2+\tau^2+x_\perp^2}\right]\,, \qquad
u^\perp = \sinh \left[\tanh^{-1}\frac{2\tau x_\perp}{L^2+\tau^2+x_\perp^2}\right]\,, \label{gub}
\eeq
and $u^\zeta=u^\phi=0$.  The parameter $L$ is the characteristic length scale of the system. In heavy-ion collisions, it is roughly the transverse size of the colliding nuclei.
 Eq.~(\ref{gub}) is called Gubser flow \cite{Gubser:2010ze,Gubser:2010ui} expressed in the coordinate system
 \beq
 ds^2=-d\tau^2 + \tau^2 d\zeta^2 + dx_\perp^2 + x_\perp^2 d\phi^2\,, \label{mink}
 \eeq
 where $\tau=\sqrt{t^2-x_3^2}$ is the proper time, $\zeta=\tanh^{-1}\frac{x_3}{t}$ is the spacetime rapidity and $x_\perp=\sqrt{x_1^2+x_2^2}$ is the transverse coordinate. The condition $u^\zeta=0$ means that the flow is boost invariant along the beam ($x_3$) direction.

 Gubser flow takes a very simple form in a cleverly chosen coordinate system $\hat{x}^\mu$ which is related to the Minkowski coordinates via a Weyl rescaling of the metric.
\beq
d\hat{s}^2=\frac{ds^2}{\tau^2}=-d\rho^2+ \cosh^2\rho (d\Theta^2+\sin^2\Theta d\phi^2) + d\zeta^2\,, \label{weyl}
\eeq
where
\beq
\sinh\rho=-\frac{L^2-\tau^2+x_\perp^2}{2L\tau}\,, \qquad \tan\Theta=\frac{2Lx_\perp}{L^2+\tau^2-x_\perp^2}\,. \label{rho}
\eeq
 In this coordinate system, the flow velocity is simply $\hat{u}^\mu=\delta^\mu_\rho$. In addition to the boost invariance, the flow respects the $O(3)$ symmetry with respect to the `polar' angles $(\Theta,\phi)$. Variables in this coordinate system will be denoted with a `hat', e.g., $\hat{u}^\mu$, $\hat{\varepsilon}$.

\section{Inviscid case}
\label{sec:3}

In this section, we solve the hydrodynamic equations (\ref{current}) in the ideal case $\eta=\kappa_i=0$. We then deform the solution in the azimuthal direction $\phi$ and compute flow harmonics $v_n$.

\subsection{Isotropic ideal solution}
 The isotropic solution (i.e., independent of $\phi$) has been obtained already in \cite{Gubser:2010ze,Gubser:2010ui} in the presence of a current $J^\mu=nu^\mu$. Assuming that all the quantities depend only on $\rho$, we can readily solve the hydrodynamic equations for $\hat{\epsilon}_0$ and $\hat{n}_0$ in the coordinates (\ref{weyl}). We then perform the Weyl transformation back to the Minkowski space $\varepsilon_0 = \hat{\varepsilon}_0/\tau^4$, $n_0=\hat{n}_0/\tau^3$ to get
  \beq
 \varepsilon_0=T^4_0f\left(\frac{\mu_0}{T_0}\right)&\propto& \frac{1}{\tau^4(\cosh\rho)^{8/3}}\,, \label{ee}
 \eeq
 \beq
 n_{0}=\mu_{0} T_0^2g\left(\frac{\mu_{0}}{T_0}\right)\propto \frac{1}{\tau^3\cosh^2\rho}\,, &\qquad&   \tilde{n}_{0}=\tilde{\mu}_{0} T_0^2\tilde{g}\left(\frac{\mu_{0}}{T_0}\right)\propto \frac{1}{\tau^3\cosh^2\rho}\,.
\eeq
These equations can be solved for $T_0$ and $\mu_0$.  It is consistent to look for the solution where $T_0$ and $\mu_0$ have the same $\rho$-dependence such that the ratios $\alpha\equiv \mu_0/T_0$, $\tilde{\alpha}\equiv \tilde{\mu}_0/T_0$ are independent of $\rho$.
We find
  \beq
 T_0=\frac{C}{\tau(\cosh\rho)^{2/3}}\,, \qquad
 \mu_0 =\frac{\alpha C}{\tau(\cosh\rho)^{2/3}}\,, \qquad
 \tilde{\mu}_0=\frac{\tilde{\alpha} C}{\tau(\cosh\rho)^{2/3}}\,,
 \label{tmu}
 \eeq
and therefore,
 \beq
 \varepsilon_0= \frac{f(\alpha)C^4}{\tau^4(\cosh\rho)^{8/3}}\,, \qquad
 n_0=\frac{\alpha g(\alpha)C^3}{\tau^3\cosh^2\rho} \,, \qquad
 \tilde{n}_0=\frac{\tilde{\alpha} \tilde{g}(\alpha)C^3}{\tau^3\cosh^2\rho}\,.
 \label{idealsol}
 \eeq
 The parameter $C$ is related to the particle multiplicity to be extracted from the experimental data.  For a massless particle species $i$ (`pion'), the relation is \cite{Gubser:2010ze,Hatta:2014jva}
   \beq
   \frac{dN_i}{dY}\approx g_i\frac{4C^3}{\pi} \,, \label{C}
  \eeq
   where $Y$ is the momentum rapidity and $g_i$ is the degeneracy factor.

\subsection{Anisotropic ideal solution}

We now perturb the solution anisotropically to introduce the $\cos n\phi$ dependence. In doing so, we shall focus on the early time regime $\tau\ll L$ (or $\rho\to -\infty$, see (\ref{rho})). As observed in \cite{Hatta:2014jva}, in this regime the perturbed solution is fully under analytical control including the viscous case to be discussed in the next section.

 Following \cite{Gubser:2010ui}, we consider the following deformation of the isotropic solution
\beq
\hat{\varepsilon}_0 \to \hat{\varepsilon}&=& \hat{\varepsilon}_0(1-\epsilon_n {\mathcal A}\delta)^4\,, \nn
\hat{n}_0\to \hat{n}&=&\hat{n}_0(1-\epsilon_n {\mathcal A}\delta')^3 \,, \nn
 (\hat{u}^\rho,\hat{u}^\Theta, \hat{u}^\phi, \hat{u}^\zeta)=(1,0,0,0)&\to& \left(1, -\epsilon_n \nu_s \hat{g}^{\Theta\Theta}\partial_\Theta {\mathcal A}, -\epsilon_n\nu_s \hat{g}^{\phi\phi} \partial_\phi  {\mathcal A},0 \right) \,, \label{an}
\eeq
 where
 \beq
  {\mathcal A} \equiv \left(\frac{2Lx_\perp}{L^2+x_\perp^2}\right)^n \cos n\phi\,,
 \eeq
 is proportional to the spherical harmonics $Y_{n,n}(\Theta,\phi) + Y_{n,-n}(\Theta,\phi)$ in the early time regime $\tau \ll L$. Note that we preserve boost invariance $u^\zeta=0$ in this paper, but the case $u^\zeta \neq 0$ was also considered in \cite{Gubser:2010ui}.
    $\epsilon_n$ is the eccentricity\footnote{In a conformal theory, the definition of eccentricity requires some care. We use \cite{Gubser:2010ui,Hatta:2014jva}
    \beq
    \epsilon_n=-\frac{\int d^2x_\perp \varepsilon^{3/4}\frac{x_\perp^n}{(L^2+x_\perp^2)^{n-1}}\cos n\phi}{\int d^2x_\perp \varepsilon^{3/4}\frac{x_\perp^n}{(L^2+x_\perp^2)^{n-1}}}\,. \label{epdef}
    \eeq

    . } which we assume to be small $\epsilon_n\ll 1$ and keep only linear terms in $\epsilon_n$.  $\delta(\rho), \delta'(\rho)$ and  $\nu_s(\rho)$ have to be determined by solving the hydrodynamic equations linearized around the isotropic solution.
 Plugging (\ref{an})  into (\ref{current}), we find the following equation for $\delta'$
 \beq
 \partial_\rho \delta' =\frac{\nu_s}{3\cosh^2\rho}n(n+1)\,. \label{delta}
 \eeq
 This turns out to be exactly the same as the equation satisfied by $\delta$  \cite{Gubser:2010ui}. Therefore, in the ideal case we have $\delta=\delta'$, which means that $T_0$ and $\mu_0$ are rescaled by the same factor $T= T_0(1-\epsilon_n {\mathcal A}\delta)$, $\mu= \mu_0(1-\epsilon_n {\mathcal A}\delta)$ and $\tilde{\mu}= \tilde{\mu}_0(1-\epsilon_n {\mathcal A}\delta)$. The ratios $\mu/T=\mu_0/T_0=\alpha$ and $\tilde{\mu}/T=\tilde{\mu}_0/T_0=\tilde{\alpha}$ are thus unchanged. At early times $\rho\to -\infty$, the right hand side of (\ref{delta}) is negligible and we can set $\delta= 1$ \cite{Hatta:2014jva}.

\subsection{$v_n$ at finite $\mu$}
\label{fini}

 In order to compute flow harmonics $v_n$, we use the Cooper-Frye formula \cite{Cooper:1974mv}
\beq
\!\!(2\pi)^3 \frac{dN_i}{dY p_T dp_T d\phi_p}\! = g_i\!\int_\Sigma (-p^\mu d\sigma_\mu) \! \left(\exp\left(\frac{u \cdot p + k \mu_i}{T}\right) + \delta f \right)\! \propto 1+2v_n(p_T)\cos n\phi_p\,, \label{cooper}
\eeq
 where we assumed the Boltzmann distribution and $\delta f$ is the deviation from the equilibrium distribution. The use of the Boltzmann distribution (rather than the Fermi/Bose distributions) may be justified for the purpose of computing the integrated $v_n$ \cite{Hatta:2014jva}.
 $\mu_i$ generically represents a set of chemical potentials for net baryon number, isospin and strangeness. We assign $k=\pm 1$ for particles with positive/negative quantum numbers mentioned above, and $k=0$ for neutral particles with respect to the corresponding quantum number. In principle, since we are assuming conformal symmetry, the formula (\ref{cooper}) should be used only for massless particles, or particles that can be approximately treated as massless (i.e., pions). However, for the sake of discussion in Section~\ref{dif}, we shall later introduce massive particles and compute their $v_n$ in the `probe approximation', namely, by neglecting their backreaction to the flow velocity. Since we add in particles in the final state that do not exist in the fluid,  the total energy is not conserved at freezeout. But the fraction of the change $\delta \varepsilon/\varepsilon \sim e^{-m/T}$ is exponentially suppressed by the mass $m$ and will be neglected.

The integral in (\ref{cooper}) is taken  over the hypersurface $\Sigma$ of constant energy density where the kinetic freezeout occurs. In the ideal case, constant $\varepsilon$ means constant $T$ since $\alpha=\mu/T$ is a constant.  Let us write the condition of constant energy density as
 \beq
 \varepsilon(\tau,x_\perp,\phi)=T^4f(\alpha)\equiv  \frac{C^4B^4}{(2L)^4}f\left(\alpha\right)=\varepsilon_c\,. \label{kinetic}
 \eeq
 Typically,  $\varepsilon_c$ is of the order of the critical energy density of the QCD phase transition. We take $\varepsilon_c=1\, \mbox{GeV}/\mbox{fm}^3$ in this paper.
Following \cite{Hatta:2014jva}, we assume that the condition (\ref{kinetic}) is reached  within the  early time regime $\tau\ll L$ where we can use the approximate solution (\ref{an}). The parameter $B$ in (\ref{kinetic}) is then related to the (position-dependent) freezeout time $\tau_f$ as
 \beq
 \tau_f(x_\perp,\phi)=\frac{(2L)^5}{B^3(L^2+x_\perp^2)^2} \left(1-3\epsilon_n \left(\frac{2Lx_\perp}{L^2+x_\perp^2}\right)^n \cos n\phi \right)\,.
 \eeq
 For consistency with our early freezeout scenario, we must have  $B^3 \gg 1$.

Under these assumptions, the integral (\ref{cooper}) can be performed analytically and the integrated $v_n$ is obtained from the formula
\beq
v_n=\frac{\int dp_T v_n(p_T)\frac{dN}{dYdp_T}}{\int dp_T \frac{dN}{dYdp_T}}\,. \label{ra}
\eeq
  In the ideal case $\delta f=0$, $v_n$ does not depend on $k$ since the factor $e^{k\mu/T}=e^{k\alpha}$ cancels in the ratio (\ref{ra}). The result  is  \cite{Hatta:2014jva}\footnote{See (61) of \cite{Hatta:2014jva}. We have corrected a mistake by a factor of 2 in the overall normalization. The same comment applies to (\ref{vn}) below.}
 \beq
 \frac{v_n}{\epsilon_n} =  \frac{9}{64}\frac{\Gamma(3n)}{\Gamma(4n)}\left(\frac{128}{B^3}\right)^n \Gamma^2\left(\frac{n}{2}\right) \frac{n^2(3n+2)^2(n-1)}{2(4n+1)} \sim B^{-3n} \propto \left(\frac{f^{3/4}}{\varepsilon_c^{3/4}L^3}\frac{dN}{dY}\right)^n\,. \label{vnn}
 \eeq
 This determines the $\alpha=\mu/T$ dependence of $v_n$. Quite generally, $f(\alpha)$ is an increasing function $\alpha$. On the other hand, $dN/dY$ is a decreasing function of $\alpha$. We shall see that, in heavy-ion collisions, the latter dependence is stronger, and as a result (\ref{vnn}) is a decreasing function of $\alpha$, or equivalently, an increasing function of the collision energy $\sqrt{s}$. Incidentally, we note that the directed flow $v_{n=1}$ vanishes, consistently with our assumption of boost-invariance.

\section{Viscous case}
\label{sec:4}

 We now turn to the viscous case $\eta, \kappa_i \neq 0$. Although the system is out of equilibrium, from the Landau matching condition we can define the local $T$ and $\mu$ using the same relations as in equilibrium
\beq
\varepsilon=T^4f\left(\frac{\mu}{T}\right)\,, \qquad
n = \mu T^2 g\left(\frac{\mu}{T}\right)\,, \qquad  \tilde{n}=\tilde{\mu} T^2 \tilde{g}\left(\frac{\mu}{T}\right)\,, \label{ns}
\eeq
 but now $\mu/T$ cannot be a constant.

\subsection{Isotropic viscous solution}

 First consider the isotropic case $\hat{u}^\mu=\delta^\mu_\rho$. Although $\mu/T$ in (\ref{charge}) is not a constant anymore, it depends only on $\rho$ (see below). Then we still  have $\hat{J}^\mu=\hat{n}\delta^\mu_\rho$  so that
\beq
n \propto \frac{1}{\tau^3 \cosh^2\rho}\,,
\eeq
 is the same as in the ideal case \cite{Gubser:2010ui}. However, the solution of the Navier-Stokes equation $\varepsilon_{NS}$ has an extra $\rho$-dependence proportional to the shear viscosity $\eta$.
In the case of vanishing chemical potentials, this $\rho$-dependence can be obtained  exactly \cite{Gubser:2010ze}
 \beq
\varepsilon_{NS} =\frac{1}{\tau^4} \frac{fC^4}{(\cosh\rho)^{8/3}}\left[1+\frac{\hat{\eta}}{9f^{1/4}C}\sinh^3\rho \ _2
F_1\left(\frac{3}{2},\frac{7}{6},\frac{5}{2};-\sinh^2\rho\right)\right]^4\,,
\eeq
where  $\hat{\eta}\equiv \eta/\varepsilon^{3/4}_{NS}$ is independent of $\rho$.

However, at finite density, $\hat{\eta}$ will depend on $\rho$, and this makes it difficult to find an exact solution. Related to this, $\eta$ can now depend on both $\varepsilon$ and $n$, and this relation can be model-dependent. We can get around this problem by assuming that $\eta$ is small.
Specifically, we rescale $\eta$ by the entropy density $s$
\beq
\eta=\bar{\eta}\left(\frac{\mu}{T}\right) s\,, \label{eta}
\eeq
 as is often done in hydrodynamic simulations. We then regard $\bar{\eta}$ as a small parameter ($\bar{\eta}\sim{\mathcal O}(10^{-1})$)   and keep only terms linear in $\bar{\eta}$. In this approximation, we may replace $\mu/T$ and $s$ in (\ref{eta}) by their equilibrium values at $\eta=0$, namely, $\mu/T=\alpha$  and
 \beq
 s\approx \frac{1}{T_0}\left(\varepsilon_0+p_0-\mu_0 n_0\right) = \frac{C^3}{\tau^3\cosh^2\rho}\left(\frac{4}{3}f(\alpha)-\alpha^2 g(\alpha)\right)\equiv \frac{C^3 }{\tau^3\cosh^2\rho}h(\alpha)\,.
 \eeq
We then find the solution valid to ${\mathcal O}(\bar{\eta})$
\beq
\varepsilon_{NS}&=&T^4f\left(\frac{\mu}{T}\right) \approx\frac{f(\alpha)C^4}{\tau^4 (\cosh\rho)^{8/3}}\left[1 + \frac{4h(\alpha)\bar{\eta}(\alpha)}{9f(\alpha)C} \sinh^3\rho \ _2F_1\left(\frac{3}{2},\frac{7}{6},\frac{5}{2};-\sinh^2\rho\right) \right] \nn
&&\qquad \qquad  \approx \frac{f(\alpha)C^4}{\tau^4 (\cosh\rho)^{8/3}}\left[1 - \frac{2h(\alpha)\bar{\eta}(\alpha)}{f(\alpha)C} \left(\frac{e^{-\rho}}{2}\right)^{2/3} \right]\,, \label{na} \\
n_{NS} &=& \mu T^2 g\left(\frac{\mu}{T}\right)  =\frac{\alpha g(\alpha)C^3}{\tau^3\cosh^2\rho}\,, \label{na2}
\eeq
 where in the second line of (\ref{na})  we focus on  the early-time regime where $\rho$ is negative and large.\footnote{The viscous Gubser solution is known to become  unphysical (the temperature becomes negative) as $\rho \to -\infty$ \cite{Gubser:2010ze}. Physically, this corresponds to  very early times and/or very large values of $x_\perp$. Our results are not sensitive to these regions. We can simply choose the initial time of the evolution to be small, but not too small.  Besides, all the $x_\perp$-integrals to be performed below are fully convergent at $x_\perp\to \infty$. }

Using (\ref{na}) and (\ref{na2}), we can eliminate $C$
\beq
\frac{\left(\frac{\mu}{T}g\left(\frac{\mu}{T}\right)\right)^{4/3}}
{f\left(\frac{\mu}{T}\right)} = \frac{(\alpha g(\alpha))^{4/3}}{f(\alpha) \left( 1  - \frac{2h(\alpha)\bar{\eta}(\alpha)}{f(\alpha)C} \left(\frac{e^{-\rho}}{2}\right)^{2/3} \right)}\,.
\eeq
Writing
\beq
\frac{\mu}{T}=\alpha + \delta \alpha(\rho)\,, \label{alpha}
\eeq
we find the deviation from constancy due to the viscosity
\beq
\delta \alpha(\rho)
\approx \frac{2h\bar{\eta}}{Cf} \frac{\left(\frac{e^{-\rho}}{2}\right)^{2/3} }{\frac{4}{3\alpha} + \frac{4g'}{3g}-\frac{f'}{f}} = \gamma \frac{ h\bar{\eta}}{Cf} \left(\frac{L^2+x_\perp^2}{2L\tau}\right)^{2/3}\,, \label{del}
\eeq
where
\beq
\gamma(\alpha)\equiv \frac{2}{\frac{4}{3\alpha} + \frac{4g'}{3g}-\frac{f'}{f}}\,.
\label{gamma}
\eeq
Note that $\gamma(\alpha) \propto \alpha$ as $\alpha\to 0$. At the freezeout time $\tau=\tau_f$,
we have the relation
 \beq
  \frac{\mu}{T} =\alpha+\delta\alpha|_{freezeout}
 \approx \alpha+ \gamma K\frac{(L^2+x_\perp^2)^2}{(2L)^4}\,.
 \label{def}
 \eeq
Finally, we can solve for $T$ and $\mu$ using (\ref{alpha}). The result is
\beq
T=\frac{C}{\tau(\cosh\rho)^{2/3}}\left(1-\frac{\delta \alpha}{3}\left(\frac{1}{\alpha}+\frac{g'}{g}\right)\right)\,,
\quad \
\mu=\frac{\alpha C}{\tau(\cosh\rho)^{2/3}} \left( 1+\frac{\delta\alpha}{3} \left(\frac{2}{\alpha}-\frac{g'}{g}\right)  \right)\,. \label{int}
\eeq

\subsection{Anisotropic viscous solution}
\label{conduct}

We now perturb the solution as in (\ref{an}). First consider the current in (\ref{charge}). $\mu/T$ now depends not only on $\rho$, but also on $\Theta$ and $\phi$. However, the dependence is of order $\eta$. (See (\ref{al}) below. Remember that for the ideal solution $\mu/T$ is constant even in the anisotropic case.) Therefore, if we neglect terms of order ${\mathcal O}(\kappa \eta \epsilon_n)$, we can approximate
$\hat{J}^\mu \approx \hat{n}\hat{u}^\mu$. Then (\ref{delta}) is still valid and we get
\beq
n=\mu T^2 g\left(\frac{\mu}{T}\right)  \approx \frac{\alpha g(\alpha)C^3}{\tau^3\cosh^2\rho}(1-\epsilon_n {\mathcal A})^3\,. \label{char}
\eeq
As for the energy density, we find
\beq
\varepsilon=T^4f\left(\frac{\mu}{T}\right) \approx \frac{f(\alpha)C^4}{\tau^4 (\cosh\rho)^{8/3}}\left(1 - \frac{2h(\alpha)\bar{\eta}(\alpha)}{f(\alpha)C} \left(\frac{e^{-\rho}}{2}\right)^{2/3}  \right) (1-\epsilon_n {\mathcal A}\delta)^4\,,
\eeq
 where \cite{Hatta:2014jva}
 \beq
 \delta \approx 1  + \frac{h(\alpha)\bar{\eta}(\alpha)}{2f(\alpha)C} \left(\frac{e^{-\rho}}{2}\right)^{2/3}\,.
 \eeq
From the constant energy condition
 \beq
 \varepsilon=\frac{C^4B^4}{(2L)^4}f(\alpha)=\varepsilon_c\,, \label{cons}
 \eeq
  we can determine the freezeout surface $\tau(x_\perp,\phi)$  in the viscous case \cite{Hatta:2014jva}
    \beq
 \tau_f(x_\perp,\phi)=\frac{(2L)^5}{B^3(L^2+x_\perp^2)^2} \left(1-\frac{3K(L^2+x_\perp^2)^2}{2(2L)^4}-3\epsilon_n \left(\frac{2Lx_\perp}{L^2+x_\perp^2}\right)^n \cos n\phi \right)\,, \label{freeze}
 \eeq
 where the `Knudsen number' is proportional to the shear viscosity
  \beq
  K = \frac{h(\alpha)\bar{\eta}(\alpha)B^2}{f(\alpha)C}\,.
\label{kn}  \eeq
(\ref{alpha}) and (\ref{int}) are also modified as $\frac{\mu}{T}=\alpha + \delta \alpha'$ where
\beq
\delta \alpha'=\delta \alpha (1+\epsilon_n {\mathcal A}) = \delta \alpha \left(1+\epsilon_n
\left(\frac{2Lx_\perp}{L^2+x_\perp^2}\right)^n \cos n\phi \right)\,, \label{al}
\eeq
 and
\beq
T&=&\frac{C}{\tau(\cosh\rho)^{2/3}}\left[1-\epsilon_n {\mathcal A}-\frac{\delta \alpha}{3}\left(\frac{1}{\alpha}+\frac{g'}{g}\right)
\right] \label{totalt} \,, \nn
\mu&=&\frac{\alpha C}{\tau(\cosh\rho)^{2/3}}\left[ 1-\epsilon_n {\mathcal A} +\frac{\delta\alpha}{3} \left(\frac{2}{\alpha}-\frac{g'}{g}\right)  \right]\,.
\eeq

\subsection{$v_n$ at finite $\mu$ and $\eta$}
\label{new}

The computation of $v_n$ is more complicated than the $\mu=0$ case. This is because $\varepsilon=const$ does not mean $T=const$, and therefore one cannot treat $T$ in the Boltzmann factor (\ref{cooper}) as a constant when integrating over the hypersurface of constant energy. In order to cope with this, we write (\ref{totalt}) as
\beq
T=T_c -\frac{T_0f'}{4f}\delta \alpha\,,
\eeq
 where
 \beq
 T_c\equiv \frac{CB}{2L} = T_0\left(1-\epsilon_n {\mathcal A} -\frac{h\bar{\eta}}{2fC}\left(\frac{L^2+x_\perp^2}{2L\tau}\right)^{2/3} \right)\,,
  \eeq
  is  constant by virtue of (\ref{cons}). We then expand the Boltzmann factor as\footnote{In this subsection we set $\tilde{\mu}=0$. The case with $\tilde{\mu}\neq 0$ will be treated in the next subsection.}
\beq
\exp\left(\frac{u \cdot p + k \mu}{T}\right)
&\approx& e^{k\alpha}e^{u\cdot p/T_c}
\exp\left[\delta \alpha \left(\frac{u\cdot p}{T_c^2}\frac{T_0f'}{4f} +k(1+\epsilon_n {\mathcal A}) \right) \right] \nn
&\approx&  e^{k\alpha}e^{u\cdot p/T_c} \left\{1+\delta \alpha \left(\frac{u\cdot p}{T_c}
 \frac{f'}{4f}  +k \right)(1+\epsilon_n {\mathcal A}) \right\} \,, \label{comp}
\eeq
 where we approximated  $T_c\approx T_0(1-\epsilon_n {\mathcal A})$ in the ${\mathcal O}(\delta \alpha)$ term.

The first term in (\ref{comp}), proportional to unity, gives the same result as in \cite{Hatta:2014jva}\footnote{For simplicity, here we ignore the contribution from the nonequilibrium part $\delta f$ in (\ref{cooper}). This has been computed in \cite{Hatta:2014jva} for a particular choice of $\delta f$. However, its $n$-dependence is strongly affected by the choice of $\delta f$ which is not unique \cite{Dusling:2009df}. Moreover, even the overall sign of this contribution is sensitive to the $p_T$-cutoff.}
 \beq
 \frac{v_n}{\epsilon_n} =  \frac{9}{64}\frac{\Gamma(3n)}{\Gamma(4n)}\left(\frac{128}{B^3}\right)^n \Gamma^2\left(\frac{n}{2}\right) \left\{\frac{n^2(3n+2)^2(n-1)}{2(4n+1)}
 -\frac{3n^3(n-1)K}{16(3n-1)}(3n^2+3n+2)\right\}\,. \label{vn}
 \eeq
 Note that $v_n/v_n^{ideal}=1-{\mathcal O}(nK)$ for $n\gg 1$ (see, however, \cite{Staig:2011wj}).
The second term in (\ref{comp}) leads to a new order ${\mathcal O}(K)$ contribution to $v_n$.  To compute it, we  borrow some results from \cite{Hatta:2014jva}.  First, the perturbed flow velocity $u^\mu$ on the freezeout surface has the following components in the coordinates (\ref{mink})
\beq
u_\perp=u_{0\perp}+\delta u_\perp \epsilon_n \cos n\phi\,, \qquad u_\phi=\delta u_\phi \epsilon_n \sin n\phi\,,
\eeq
 where
\beq
&&u_{\perp 0}=2x_\perp \frac{(2L)^5}{B^3(L^2+x_\perp^2)^3}\,, \nn
&&\delta u_\perp = \frac{3(2L)^5}{B^3(L^2+x_\perp^2)^4}\left(\frac{2Lx_\perp}{L^2+x_\perp^2}\right)^{n-1}
L(n(L^2-x_\perp^2)-4x_\perp^2)\,, \nn
&&\delta u_\phi=-\frac{3n}{2}\frac{(2L)^5}{B^3(L^2+x_\perp^2)^2}
\left(\frac{2Lx_\perp}{L^2+x_\perp^2}\right)^n\,.
\eeq
 (The viscosity can be neglected here.) The exponential factor in the Boltzmann distribution reads
\beq
\frac{p\cdot u}{T_c} = \frac{1}{T_c}\left[-m_T\cosh(\zeta-Y)+p_T u_\perp \cos (\phi-\phi_p) -\frac{p_Tu_\phi}{x_\perp}\sin(\phi-\phi_p)\right] \equiv U+\epsilon_n\delta U\,, \label{du}
\eeq
 where $m_T=\sqrt{m^2+p_T^2}$ is the transverse mass.
The volume element of the constant energy hypersurface is
\beq
-p^\mu d\sigma_\mu = x_\perp \tau_f \left(m_T\cosh(\zeta-Y)-p_T\cos(\phi-\phi_p)\frac{\partial \tau_f}{\partial x_\perp} + \frac{p_T}{x_\perp}\sin(\phi-\phi_p)\frac{\partial \tau_f}{\partial \phi}\right)d\zeta dx_\perp d\phi\,, \nn
\label{sur}
\eeq
 where $\tau_f$ is given by (\ref{freeze}) with the viscous term set to zero.
 Finally, we need the more precise version of (\ref{def})
 \beq
 \delta \alpha
 \approx \gamma K\frac{(L^2+x_\perp^2)^2}{(2L)^4} (1+2\epsilon_n {\mathcal A})\,.
 \eeq

Armed with these formulas, let us decompose the contribution from the second term in (\ref{comp}) as
\beq
(2\pi)^3 \frac{dN}{dY p_T dp_T d\phi_p} &\sim& e^{k\alpha}\int_\Sigma (-p^\mu d\sigma_\mu)  e^{U+\epsilon_n \delta U} \delta \alpha \left((U+\epsilon_n \delta U)
 \frac{f'}{4f}  +k \right)(1+\epsilon_n {\mathcal A}) \nn
 &\equiv& (\delta J_1+\delta J_2+\delta J_3)\epsilon_n \cos n\phi_p \,, \label{j123}
\eeq
corresponding to the three terms in (\ref{sur}). Consider $\delta J_1$ first.  To ${\mathcal O}(\epsilon_n)$ we have to evaluate
\beq
\delta J_1\sim \epsilon_n \frac{2L\gamma K e^{k\alpha}}{B^3}m_T\int dx_\perp x_\perp
\int d\zeta d\phi \cosh(\zeta-Y)e^U \delta U \left(\frac{f'}{4f}U +\frac{f'}{4f} + k\right)\,.
\eeq
 This can be efficiently evaluated using the trick introduced in \cite{Hatta:2014jva} (see Eq.~(73) there). The $\phi$-integral gives Bessel functions $I_n(z)$
  where
 \beq
 z\equiv \frac{p_Tu_{\perp 0}}{T_c}=\frac{2x_\perp p_T(2L)^5}{T_cB^3(L^2+x_\perp^2)^3}\,.
 \eeq
 This can be expanded as $I_n(z)\sim z^n$ anticipating that the subsequent $p_T$-integral is dominated by the region $z< 1$. We thus find
\beq
\delta J_1 &\approx&  \frac{2L\gamma K  e^{k\alpha}}{B^3}m_T\int_0^\infty dx_\perp x_\perp \frac{4\pi  z^{n}}{2^n(n-1)!}\frac{1}{u_{\perp 0}}\left(\delta u_\perp -\frac{\delta u_\phi}{x_\perp}\right) \nn
 &&\times \left[\frac{f'}{4f}\left(nK_1(m_T/T_c) -\frac{m_T}{2T}\left(K_0(m_T/T_c)+K_2(m_T/T_c)\right)\right) + kK_1(m_T/T_c)\right]
\nn
&=& \frac{2L\gamma K e^{k\alpha}}{B^3} 9\pi L^2m_T\left(\frac{64p_T}{T_cB^3}\right)^n\frac{n(n-1)\Gamma(3n)}{(3n-1)\Gamma(4n)}
 \nn
&&  \times \left[\frac{f'}{4f}\left(nK_1(m_T/T_c) -\frac{m_T}{2T}\left(K_0(m_T/T_c)+K_2(m_T/T_c)\right)\right) + kK_1(m_T/T_c)\right]\,.
\label{bes}
 \eeq

 The correction to $v_n$ can be calculated from the formula (cf. (\ref{ra}))
 \beq
 \delta v^1_n \equiv \frac{\int_0^\infty dp_T p_T\delta J_1}{\int_0^\infty dp_T p_T J^0}\frac{\epsilon_n}{2}\,, \label{cor}
 \eeq
  where $J^0$ is the azimuthally symmetric part (cf. Eq.~(45) of \cite{Hatta:2014jva})
  \beq
  J^0=4\pi m_T K_1(m_T/T_c)\frac{16L^3}{B^3}e^{k\alpha}\,.
  \eeq
In the massless case $m_T=p_T$, the integral can be done exactly and we find
\beq
\frac{\delta v^1_n}{\epsilon_n} =\frac{9\gamma K}{128} \left(k-\frac{3f'}{4f}\right) \left(\frac{128}{B^3}\right)^n \frac{n^2(n-1)\Gamma(3n)}{(3n-1)\Gamma(4n)} \Gamma\left(\frac{n}{2}+2\right)\Gamma\left(\frac{n}{2}\right)\,, \label{v1}
\eeq
 and from (\ref{vnn}),
\beq
\frac{\delta v^1_n}{v_n^{ideal}}=\frac{\gamma K}{4} \left(k-\frac{3f'}{4f} \right) \frac{n(n+2)(4n+1)}{(3n-1)(3n+2)^2}\,. \label{no}
\eeq
 It is important to emphasize that (\ref{no}) is induced by the combined effect of the chemical potential and the viscosity. It vanishes when $\eta=0$ or $\alpha=\mu/T=0$ because $\gamma(0)=0$ (cf. (\ref{gamma})). Compared with (\ref{vn}) which schematically reads $\delta v_n/ v_n^{ideal} \sim -nK$, we notice that (\ref{no}) is not enhanced by a factor of $n$, hence subleading at large $n$. However, it is the leading contribution to the \emph{difference} in $v_n$ between particles ($k=1$) and antiparticles ($k=-1$).  If $\mu=\mu_B>0$ is the baryon chemical potential, the protons have larger $v_n$ than the antiprotons. We shall study this effect in detail later.

In fact, for protons the approximation $m_T\approx p_T$ is not valid. Instead, we now assume $m_T\gg T$ and reevaluate $\delta v_n$. Note that when $m_T\gg T$, $\delta J_1$ is parametrically larger than $\delta J_{2,3}$, so it is  enough to consider only $\delta J_1$.

When $m_T\gg T$, the Bessel function is independent of the order
\beq
K_i(m_T/T_c)\approx \sqrt{\frac{\pi T_c}{2m_T}} e^{-m_T/T_c}\,,
\eeq
 so that (\ref{bes}) becomes
 \beq
 \delta J_1\approx  \frac{2\gamma K e^{k\alpha}}{B^3} 9\pi L^3 m_T\left(\frac{64p_T}{T_cB^3}\right)^n\frac{n(n-1)\Gamma(3n)}{(3n-1)\Gamma(4n)}K_1(m_T/T_c)
 \left[\frac{f'}{4f}\left(n -\frac{m_T}{T_c}\right) + k\right]\,.
 \eeq
On the other hand, from Eq.~(47) of \cite{Hatta:2014jva},
\beq
\delta J^{ideal}_1 =4\pi e^{k\alpha}\frac{m_T}{B^3} K_1(m_T/T_c)\frac{\Gamma(3n)}{ \Gamma(4n)}
9 L^3 \left(\frac{64p_T}{T_cB^3}\right)^n   (n-1)
  \frac{2(3n+2)}{4n+1}  \,, \label{deltaJ1}
  \eeq
 The $p_T$-integral can be evaluated by the saddle point at $p_T^*=\sqrt{nmT_c}$ for $m\gg nT_c$ and we obtain
 \beq
 \frac{\delta v_n^1}{v_n^{ideal}}\approx \frac{\gamma K}{4}\frac{n(4n+1)}{(3n-1)(3n+2)}\left( \frac{f'}{4f}\left(\frac{n}{2}-\frac{m}{T_c}\right)+k\right)\,.
 \label{pto}
 \eeq
 The $k$-independent part is order $\frac{m}{T}K\gg  nK$, but we shall see later that it is numerically small for realistic values of $m$ because the factor $\gamma f'/f$ is small. The $k$-dependent term is again of order ${\mathcal O}(K)$ without an enhancement by a factor of $n$.
\\

The evaluation of $\delta J_{2,3}$ in (\ref{j123}) can be done similarly, though it is considerably more tedious. Here we only show the final result in the massless case $m=0$, relegating the details to Appendix
\beq
\frac{\delta v_n^{2+3}}{\epsilon_n} =\frac{9\gamma K}{128} \left(k-\frac{3f'}{4f}\right) \left(\frac{128}{B^3}\right)^n \frac{n^3(n-1)\Gamma(3n)}{(3n-1)\Gamma(4n)} \Gamma\left(\frac{n}{2}+1\right)\Gamma\left(\frac{n}{2}\right)\,. \label{ap}
\eeq
Comparing with (\ref{v1}), we notice that $\delta v_n^1= \frac{n+2}{2n} \delta v_n^{2+3}$. Actually, this relation was repeatedly observed in \cite{Hatta:2014jva} when computing other contributions to $v_n$. We do not have a simple explanation for this.


Summing all the contributions including the previously computed term \cite{Hatta:2014jva}, our final result  of the viscous correction $\delta v_n$ in the massless case is
\beq
\frac{\delta v_n}{\epsilon_n}&=& \frac{K}{256}\frac{\Gamma(3n)}{\Gamma(4n)} \left(\frac{128}{B^3}\right)^n \Gamma^2\left(\frac{n}{2}\right)\frac{n^3(n-1)}{3n-1}
\nn && \qquad \times \Biggl\{-\frac{27}{4}(3n^2+3n+2) +9\gamma\left(\frac{3n}{2}+1\right)\left(k-\frac{3f'}{4f}\right) \Biggr\}\,.
\label{final}
\eeq
 The second term in the curly brackets is the new contribution at finite density. It is subleading in $n$, and actually the factor $\gamma(\alpha)$ is also numerically small. However, it gives the leading contribution to the difference in $v_n$ between particles and antiparticles.

\subsection{Isospin chemical potential}
\label{iso}

In the previous subsection, we computed $v_n$ of particles which couple to the `large' chemical potential $\mu$. Here let us compute $v_n$ of particles neutral under $\mu$ but charged under $\tilde{\mu}$. We have in mind the charged pions $\pi^{\pm}$ in the presence of the isospin chemical potential.
We start with the formula (cf. (\ref{char}))
\beq
\tilde{n}= \frac{\tilde{\alpha} \tilde{g}(\alpha)C^3}{\tau^3 \cosh^2\rho}(1-\epsilon_n {\mathcal A})^3 = \tilde{\mu}T^2\tilde{g}(\mu/T)\,.
\eeq
We treat $\tilde{\alpha}=\tilde{\mu}/T$ as a small parameter and keep only terms linear in $\tilde{\alpha}$. Dividing by $T^3$ from (\ref{totalt}) and using $\mu/T=\alpha + \delta \alpha(1+\epsilon_n{\mathcal A})$, we find
 \beq
\frac{\tilde{\mu}}{T}= \frac{\tilde{\alpha} \tilde{g}(\alpha) (1-\epsilon_n {\mathcal A})^3}{\tilde{g}(\mu/T)\left(1-\epsilon_n {\mathcal A} -\frac{\delta \alpha}{3} \left(\frac{1}{\alpha}+\frac{g'}{g}\right)\right)^3} \approx \tilde{\alpha}\left(1 + \delta \alpha \left(\frac{1}{\alpha}+\frac{g'}{g}-\frac{\tilde{g}'}{\tilde{g}}\right)(1+\epsilon_n {\mathcal A})  \right)\,.
 \eeq
  The fugacity factor thus becomes
  \beq
  e^{k\tilde{\mu}/T} \approx e^{k\tilde{\alpha}}\left(1+k\tilde{\alpha}\delta \alpha \left(\frac{1}{\alpha}+\frac{g'}{g}-\frac{\tilde{g}'}{\tilde{g}}\right)(1+\epsilon_n {\mathcal A}) \right)\,.
  \eeq
  As before, the factor $e^{k\tilde{\alpha}}$ drops out in the computation of $v_n$.
We see that the only difference from the previous case (\ref{comp}) is that $k$ is replaced by
\beq
k\frac{\tilde{\alpha}}{\alpha} \left(1+\frac{\alpha g'}{g}-\frac{\alpha \tilde{g}'}{\tilde{g}}\right)\,.
\label{rep}
\eeq
Thus the final result is the same as (\ref{final}) except that $k$ is replaced by (\ref{rep}).

\section{Phenomenological inputs}
\label{sec:5}

This section serves as a preparation for the next section where we compare our results with the experimental data.

\subsection{Models}

 In order to make quantitative predictions, we need models for the functions $f$, $g$, $\tilde{g}$ defined in (\ref{eos}). Here we  consider two extreme scenarios in terms of the interaction strength.

\subsubsection{Free quark-gluon gas}

The energy density of free, massless three flavor QCD is
\beq
\varepsilon&=&3p=\frac{8\pi^2}{15}T^4+6\sum_{q=u,d,s} \left(\frac{7\pi^2}{120}T^4+\frac{\mu^2_qT^2}{4}
+\frac{\mu^4_q}{8\pi^2} \right)\,, \nn
n_q&=& \frac{\partial p}{\partial \mu_q}= \mu_qT^2\left(1+\frac{\mu^2_q}{\pi^2 T^2}\right)\,, \label{freegas}
\eeq
 where $\mu_u=\frac{\mu_B}{3}+\frac{\mu_I}{2}$, $\mu_d=\frac{\mu_B}{3}-\frac{\mu_I}{2}$ and $\mu_s=\frac{\mu_B}{3}-\mu_S$. $\mu_B$, $\mu_I$ and $\mu_S$ are the baryon, isospin and strangeness chemical potentials, respectively. Since the net strangeness is zero in heavy-ion collisions,  we set $\mu_S=\mu_B/3$ and obtain ($\alpha=\mu_B/T$)
\beq
f(\alpha) = \frac{19}{12}\pi^2 + \frac{\alpha^2}{3} + \frac{\alpha^4}{54\pi^2}\,,
\label{freeq} \eeq
\beq
g(\alpha)= g_B(\alpha)=2\left(1+\frac{\alpha^2}{9\pi^2}\right)\,, \qquad \tilde{g}(\alpha)=g_I(\alpha)=\frac{1}{2}\left(1+\frac{\alpha^2}{3\pi^2}\right)\,. \label{freen}
 \eeq
It turns out that, due to the large denominators $9\pi^2$ or $3\pi^2$, the effect of $g$ and $\tilde{g}$ on $v_n$ is numerically small.

\subsubsection{${\mathcal N}=4$ SYM at finite ${\mathcal R}$-charge chemical potential}

Next we consider strongly coupled ${\mathcal N}=4$ supersymmetric Yang-Mills theory at finite $R$-charge chemical potential $\mu$. This theory is conformal, and in the limit of strong coupling and at large $N_c$, the thermodynamic quantities can be computed from the AdS/CFT correspondence. The results are \cite{Erdmenger:2008rm}
\beq
\varepsilon&=&3p=\frac{3\pi^2 N_c^2T^4}{8} \frac{1}{2^4}\left( \sqrt{1+\frac{2\mu^2}{3\pi^2T^2}}+1\right)^3 \left(3\sqrt{1+\frac{2\mu^2}{3\pi^2T^2}}-1\right)\,, \label{strongq}\\
n&=&\frac{\partial p}{\partial \mu}=\frac{\mu N^2_cT^2}{16} \left(\sqrt{1+\frac{2\mu^2}{3\pi^2T^2}}+1\right)^2
\,,\label{n}
\eeq
 where $n$ is the $R$-charge density. The shear viscosity is given by $\eta=\frac{s}{4\pi}$.

There are uncertainties when treating this model as a proxy of strongly coupled QCD, such as the value of $N_c$ and the proportionality constant between $\mu$ and $\mu_B$. However, in practical fits, the normalization of $f\propto N_c^2$ can be absorbed by a change in $L$ (cf. Eq.~(\ref{kinetic})). Moreover, as long as $\mu \sim {\mathcal O}(\mu_B)$, the two functions (\ref{freeq}) and (\ref{strongq}) are qualitatively not so different in shape for $\mu\sim {\mathcal O}(T)$. As a result, the quality of fits is similar in the two cases despite the huge differences in the underlying dynamics. Therefore, in the next section we show only the results based on (\ref{freeq}) and (\ref{freen}).


\subsection{Freezeout conditions}

\begin{figure}[htbp]
   \includegraphics[width=95mm]{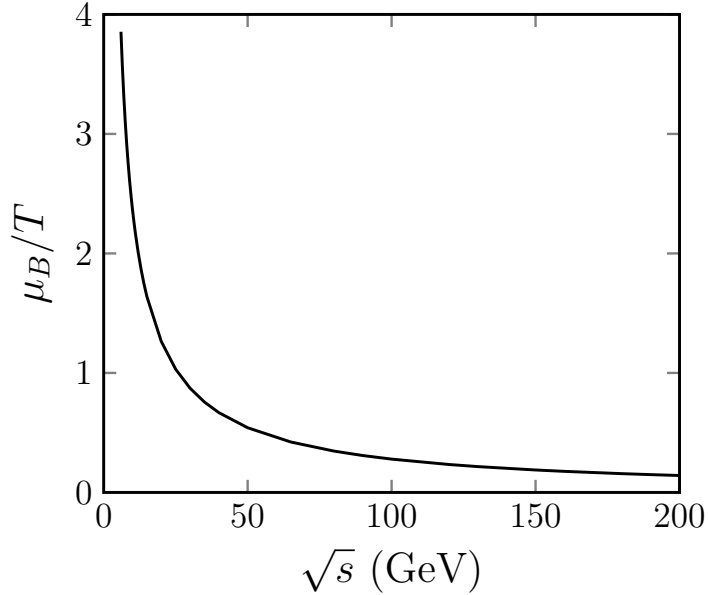}
 \caption{$\mu_B/T$ at freezeout as a function of the collision energy $\sqrt{s}$ (per nucleon).\label{fig1}}
\end{figure}

We employ the following phenomenological parametrization \cite{Cleymans:2005xv} of the  freezeout temperature $T$ and chemical potential $\mu_B$ (in units of GeV) as a function of the collision energy  $\sqrt{s}$ (per nucleon, in units of GeV)
\beq
T(\mu_B)=a-b\mu^2_B-c\mu^4_B\,, \qquad \mu_B=\frac{d}{1+e\sqrt{s}}\,, \label{red}
\eeq
with $a=0.166$, $b=0.139$, $c=0.053$, $d=1.308$, $e=0.273$. This gives $\mu_B/T$ as a function of $\sqrt{s}$ as shown in Fig.~\ref{fig1}. The curve is well approximated by $\mu_B/T \approx d/(a e \sqrt{s})\approx 29/\sqrt{s}$. Actually, $T$ and $\mu_B$ here are the \emph{chemical} freezeout parameters which are in general different from those entering the Cooper-Frye formula (\ref{cooper}) used at the \emph{kinetic} freezeout. However, in our model only the ratio $\mu_B/T$ matters, and this ratio is roughly constant as we have seen. We thus use the relation in Fig.~\ref{fig1} for the evaluation of $v_n$.

The parameter $C$ also depends on $\sqrt{s}$ via (\ref{C}). We use the following empirical formula for the charged particle multiplicity \cite{Andronic:2004tx}
\beq
\frac{dN_{ch}}{dY}\approx 2\frac{4C^3}{\pi}\approx 148 (\sqrt{s})^{0.3}\,,
\label{mul}
\eeq
 where the factor of 2 counts the degeneracy between $\pi^{+}$ and $\pi^-$.
 From  (\ref{cons}) and (\ref{mul}), we see that the Knudsen number (\ref{kn}) behaves as
  \beq
  K\approx \frac{8h\bar{\eta}L^2 \sqrt{5\varepsilon_c}}{37\pi f^{3/2}(\sqrt{s})^{0.3}} \sim \frac{\bar{\eta}L^2}{f^{1/2}(\sqrt{s})^{0.3}}\sim \frac{\bar{\eta}(\mu_B/T)^{0.3}}{\sqrt{1+\#(\mu_B/\pi T)^2}}\,,
  \label{knud}
\eeq
where $L$ is in units of fermi and $\varepsilon_c=1$ is in units of GeV/fm$^{3}$. Putting aside the potential $\mu_B$-dependence of $\bar{\eta}=\eta/s$, we see that  $K$ is an increasing function of $\mu_B$ (up to $\mu_B \lesssim \pi T$ in our model) or a decreasing function of $\sqrt{s}$.

In fact, up to the RHIC energy, we find that the following parametrization also gives a good description of the data \cite{Andronic:2014zha}
\beq
\frac{dN_{ch}}{dY}=72\ln \frac{(\sqrt{s})^2}{1.41}\,. \label{dnlog}
\eeq
 We shall also use this in Section \ref{energydep}.

\section{Comparison with the experimental data}
\label{sec:6}

In this section, we compare our  results with three different experimental data: (i) the $n$-dependence of $v_n$ measured at the LHC; (ii) the collision energy dependence of $v_2$ measured at the SPS; (iii) the difference in $v_2$ between particles and antiparticles measured at RHIC.

\subsection{Higher harmonics $v_n$}

\begin{figure}[tbp]
   \includegraphics[width=100mm]{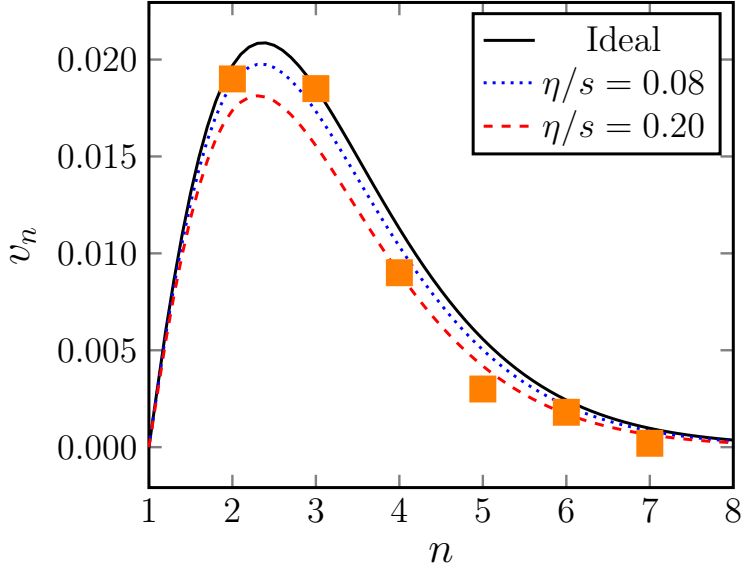}
 \caption{$v_n$ versus different values of $n$ from $2$ to $7$ measured by the CMS collaboration (0-0.2\% centrality) \cite{CMS:2012tba}. The black solid curve represents the ideal hydrodynamic result, while the blue dotted curve and red dashed curve correspond to the viscous results with $\eta/s=0.08$ and $\eta/s=0.2$, respectively. See, also, Ref.~\cite{Staig:2011wj}. \label{fig2}}
\end{figure}

The CMS collaboration at the LHC has measured the $p_T$-integrated $v_n$ in lead-lead collisions at $\sqrt{s}=2.76$ TeV up to rather high orders ($n\le 7$) \cite{CMS:2012tba}.
Using (\ref{vnn}) and (\ref{final}) together with the phenomenological inputs in the previous section, we can evaluate $v_n$ and compare with the CMS data.\footnote{The CMS uses the $p_T$ cuts $0.3<p_T<3$ GeV while our analytical result is integrated over all $p_T$. We checked that the quality of the fit is unchanged by introducing cuts in our model.}   The result is  shown in Fig.~\ref{fig2}. Here we set $\epsilon_n =0.018$ for all different values of $n$. Taking $\epsilon_n$ to be independent of $n$ may be a good approximation for the very central (0-0.2\% centrality) nucleus collisions.\footnote{The value 0.018 may seem a bit too small. This may be due to our nonstandard definition of $\epsilon_n$ (\ref{epdef}).} The parameter $L$ is set to $17\, \textrm{fm}$. The corresponding value of $B$ in (\ref{vn}) is $B^3\approx 26.7$ which is consistent with the assumption $B^3 \gg 1$.

As a matter of fact, since $\mu_B\approx 0$ at the LHC, the new term at $\mu_B>0$ (the term proportional to $\gamma$ in (\ref{final})) is negligibly small, and the present fit could have been done in \cite{Hatta:2014jva} treating $B$ as a fitting parameter. By expressing $B$ in terms of observables as we have done here, we can test our result at lower energies or higher chemical potentials $\mu_B\sim {\mathcal O}(T)$. Note that since $B^3$ is larger at lower energies, $v_n \sim e^{-n\ln (4B^3/27)}$ \cite{Hatta:2014jva} decreases faster with $n$, and this will make the measurement of higher harmonics difficult at low energies \cite{Pandit:2012mq}.

\subsection{Energy dependence of $v_2$}
\label{energydep}

\begin{figure}[htbp]
\begin{minipage}{0.45\hsize}
  \begin{center}
   \includegraphics[width=75mm]{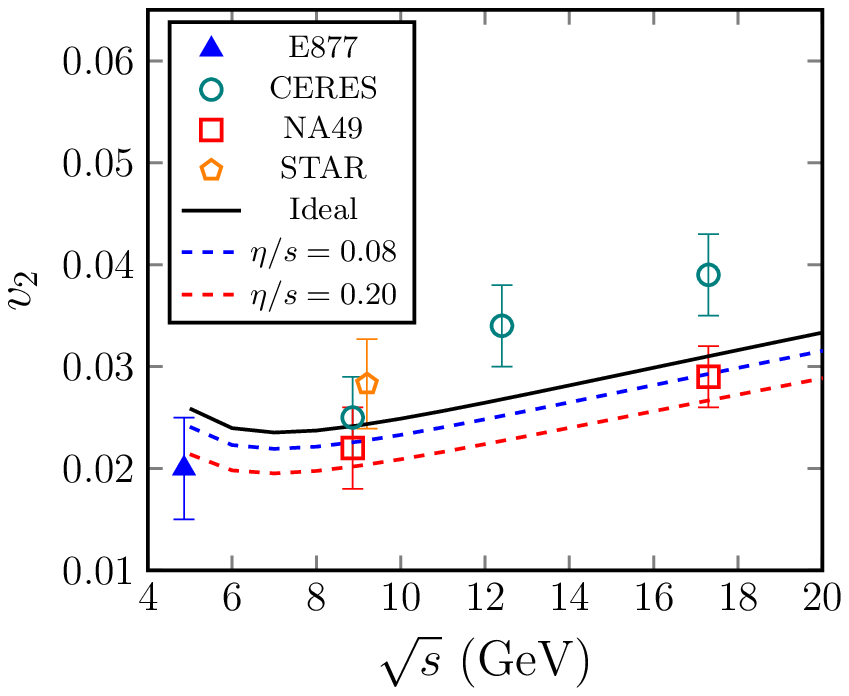}
  \end{center}
 \end{minipage}
 \begin{minipage}{0.45\hsize}
  \begin{center}
   \includegraphics[width=75mm]{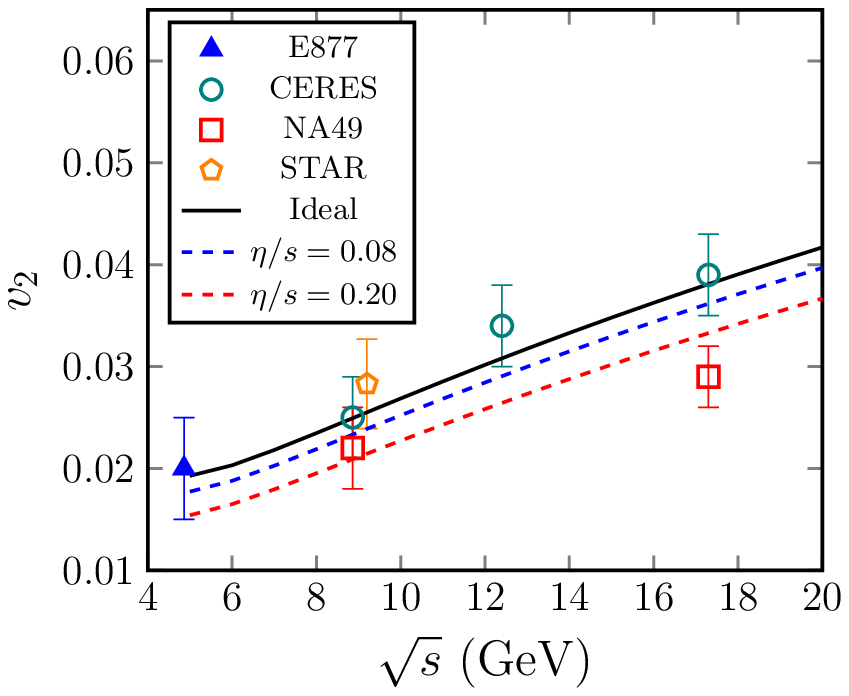}
  \end{center}
 \end{minipage}
  \caption{ The energy dependence of $v_2$ compared with the SPS data \cite{Andronic:2014zha} for mid-central collisions. We used (\ref{mul}) and (\ref{dnlog}) in the left and right plots, respectively.
 \label{fig3}}
\end{figure}

Next we turn to the energy dependence of the elliptic flow $v_{n=2}$ for which there are already a wealth of experimental data from the SPS and the RHIC BES program \cite{Alt:2003ab,Adamczyk:2012ku}.
We  compare our formulas (\ref{vn}) and (\ref{final}) (with $k=0$) for $n=2$ with the SPS, mid-central data collected in the low energy region $\sqrt{s}<20\, \textrm{GeV}$ \cite{Alt:2003ab,Andronic:2014zha}.\footnote{The SPS data do not have a low-$p_T$ cutoff while the RHIC data have $p_T>0.2\,$GeV. Our analytical formula, integrated over all $p_T$, should fare better with the SPS results. }\footnote{We thank Anton Andronic for correspondence about the SPS data.} The result with three different values of $\eta/s$ is shown in Fig.~\ref{fig3} where we tried both (\ref{mul}) and (\ref{dnlog}), the latter actually gives a better description of $dN_{ch}/dY$ in this low energy region. The other parameters are chosen as $L=15.5\,$fm and $\epsilon_2=0.32$. The value of $L$ here is slightly smaller than the one ($L=17\,$fm) used in Fig.~\ref{fig2}. This is consistent with the perception that the QGP droplet is larger at higher energies at the time of thermalization.
 The rise of $v_2$ with energy is nicely reproduced by our formula and attributed to the rise of $dN_{ch}/dY$. It turns out that the newly calculated viscous correction in Section~\ref{new} (the last term in (\ref{final})) is numerically very small (about an order of magnitude smaller than the first term in (\ref{final})) even in the highest density region.

Unfortunately, this fit, which agrees reasonably well with the low energy data, overshoots the high energy RHIC data at $\sqrt{s}=200\,\textrm{GeV}$ \cite{Esumi:2002vy,Ray:2002md} in similar centrality bins  by a factor of 2 (assuming that $\epsilon_2$ is independent of energy). This is because the rise of $dN/dY$ with energy is too steep. If we artificially reduce the exponent in (\ref{mul}) as $0.3\to 0.23$, for example, we get a decent description of $v_2$ over a  broader range in $\sqrt{s}$.\footnote{Note that (\ref{mul}) is for central collisions. The exponent may indeed be smaller for mid-central collisions.} Alternatively, the dependence $v_2\sim (dN/dY)^2$ from (\ref{vnn}) may be too strong, and the experimental data actually suggest a weaker $dN/dY$-dependence  \cite{Alt:2003ab}.  While we do not have a resolution of this problem in the present framework, it seems qualitatively correct that $v_2$ is directly proportional to the multiplicity to some positive power, and therefore it is an increasing function of $\sqrt{s}$ (see, also, Section  VII of \cite{Hatta:2014jva}).

\subsection{Difference in $v_n$ between particles and antiparticles}
\label{dif}

Finally, we investigate the difference in $v_2$ between particles and antiparticles which has been measured by the STAR collaboration at RHIC \cite{Adamczyk:2013gv,Adamczyk:2013gw} and attracted some attention from theoretical viewpoints  \cite{Burnier:2011bf,Steinheimer:2012bn,Xu:2013sta,Hongo:2013cqa}.
For a hadron with the quantum numbers $(\mathcal{B},I,S)$ (baryon number, isospin, strangeness), we assign the fugacity factor
\beq
\exp\left(\frac{\mathcal{B}\mu_B+I\mu_I+S\mu_S}{T}\right)\,.
\eeq
($S=-1$ for the strange quark.)
In heavy-ion collisions, $\mu_I<0$ since the colliding nuclei are neutron rich, and $\mu_S\approx \mu_B/3$ since the net strangeness vanishes. 
The latter condition implies that we should not treat $\mu_S$ as a small perturbation. Indeed, various estimates of $r_S\equiv \mu_S/\mu_B$ based on the SPS \cite{BraunMunzinger:1999qy} and RHIC \cite{Zhao:2014mva} data, and also from lattice QCD   \cite{Bazavov:2012vg,Bazavov:2014xya} all found similar values within the range $0.21 < r_S<0.27$.
 We thus regard $\mu_S$  as a shift of $\mu_B$ for strange hadrons and treat it as a fitting parameter, anticipating that the value of $r_S$ should come out in the window $0.2< r_S< 1/3$. On the other hand, we regard $r_I\equiv \mu_I/\mu_B$ as a small parameter compared to unity and use the result obtained in Section \ref{iso}.

Let us  define the difference in $v_n$ between hadrons $X$ and antihadrons $\bar{X}$  as
\beq
\Delta v_n^X \equiv v_n^X -v_n^{\bar{X}}\,.
\eeq
This can be evaluated from (\ref{final}) and (\ref{rep}). Focusing now on the elliptic flow case $n=2$, we can immediately write down the following `master formula'
 \beq
 \Delta v_2^X= \epsilon_2\frac{6144}{35B^6} \gamma K \left[\mathcal{B}+r_S S+r_I I \left(1+\frac{\alpha g'}{g}-\frac{\alpha g'_I}{g_I}\right)\right]\,. \label{master}
 \eeq
  By construction, (\ref{master}) has been derived for massless particles. In the massive case, we observe that the following ratio
\beq
\frac{\Delta v^X_2}{v_2^{X,ideal}} = \frac{9\gamma K}{40} \left[\mathcal{B}+r_S S+r_I I  \left(1+\frac{\alpha g'}{g}-\frac{\alpha g'_I}{g_I}\right)\right]\,, \label{from}
\eeq
is exactly independent of $m$.\footnote{As already noted in Section~\ref{fini}, we introduce massive particles in the probe approximation, namely, we let these particles flow with the same flow velocity and neglect their backreaction to the velocity. While this causes some inconsistencies such as energy nonconservation, we expect that the essential features of (\ref{from}) (the proportionality to the viscosity, $\mu$'s and the corresponding quantum numbers) are robust.} This is due to the nontrivial cancelation of $p_T$-integrals such as (\ref{cor})   in the ratio for the $k$-dependent part.   In order to get  $\Delta v_2$ itself, we must multiply (\ref{from}) by \cite{Hatta:2014jva}
\beq
 v_2^{ideal}=\epsilon_2 \frac{ 2^{11}}{35B^{6}T}
 \frac{\int dp_T\, p_T^{3}\left(\frac{m_T}{T}K_1(m_T/T)+4 K_0(m_T/T)\right)}{\int dp_T \, p_T m_T K_1(m_T/T)}\,. \label{vni}
 \eeq
  The $m$-dependence of (\ref{vni}) is sensitive to the cutoffs of the $p_T$-integral, but overall the dependence is not very strong.  For simplicity, in this study  we ignore the $m$-dependence of $v_2^{ideal}$ and use (\ref{master}) for all hadron species. It is not difficult to implement this mass effect, but there are  other subtleties which are not taken into account, either.\footnote{For instance, the STAR collaboration uses 0-80\% centrality events to measure $\Delta v_2$. This reduces the effective value of $\mu_B$ by about 20\% \cite{Adamczyk:2013gw} and partly cancels the above mass effect for baryons when computing $\Delta v_2$.  }  Clearly, it is desirable that the experimental results are plotted in the form (\ref{from}) in order to avoid various systematic uncertainties.

The most important feature of (\ref{master}) or (\ref{from}) is that $\Delta v_2$ is proportional to \emph{both } the shear viscosity $\eta$ and the chemical potentials. (Remember that $\gamma$ as defined in (\ref{gamma}) is roughly proportional to $\mu_B$.) This in particular means that $\Delta v_2^{\pi}$ can be nonzero in viscous hydrodynamics in the presence of the isospin chemical potential.

\begin{figure}[htbp]
  \begin{center}
   \includegraphics[width=125mm]{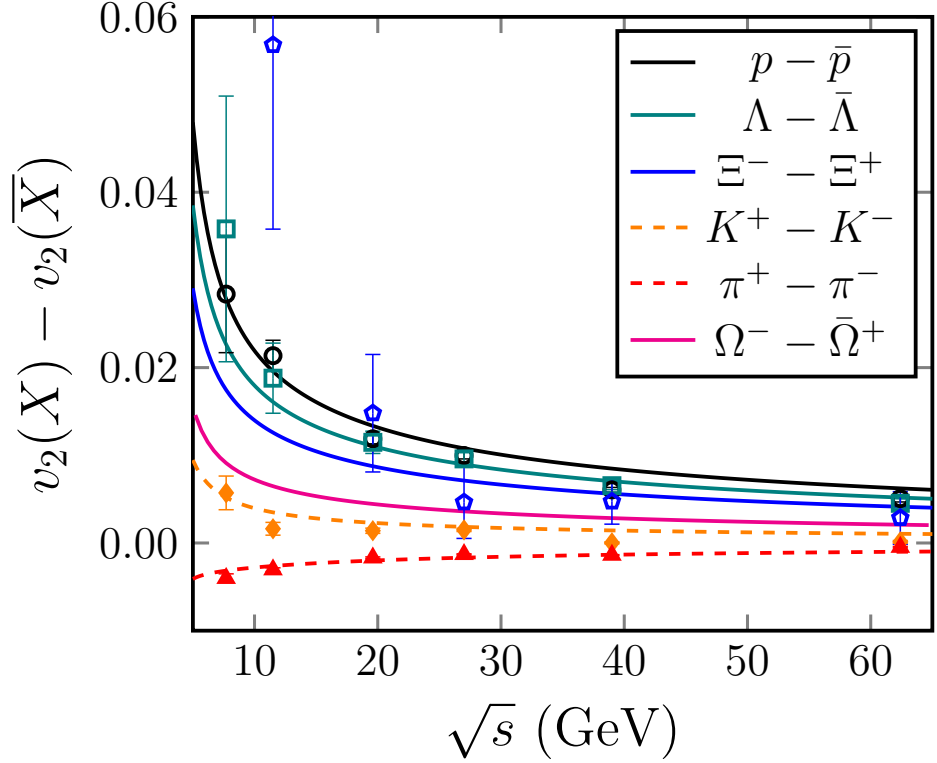}
  \end{center}
 \caption{ $\Delta v_2^{X}\equiv v_2(X)-v_2(\bar{X})$ as a function of $\sqrt{s}$  compared with the STAR data for five different species of hadrons \cite{Adamczyk:2013gv,Adamczyk:2013gw}. We use open symbols for baryons and filled symbols for mesons. The color of data points is chosen to match the color of the corresponding line for each hadron species. We have used $\eta/s=0.2$, $\mu_S=0.23\mu_B$ and $\mu_I=-0.15\mu_B$.
 \label{fig4}}
\end{figure}

  Let us confront  (\ref{master}) with the data. The STAR collaboration has measured $\Delta v_2^{X}$ for $X=\pi^+,K^+,p,\Lambda,\Xi^-$ \cite{Adamczyk:2013gv,Adamczyk:2013gw}. This is plotted in Fig.~\ref{fig4} together with our fit based on  (\ref{master}) with $\eta/s=0.2$.  We have used $L=15.5\,\textrm{fm}$ and $\epsilon_2=0.32$ as in Fig.~\ref{fig3}, and used the fit parameters $r_S=0.23$ and $r_I=-0.15$, the former is consistent with our expectation mentioned above.
  The steep rise of $\Delta v_2^{X}$ for baryons towards the low-$\sqrt{s}$ region is due to the rough proportionality $\Delta v_2 \propto \gamma \propto \mu_B$. Compared to this, the $\mu_B$-dependence of the factor $1/B^6$ is subleading. Since $p,\Lambda, \Xi^-,K^+$ have $(\mathcal{B},I,S)=(1,\frac{1}{2},0)$, $(1,0,-1)$, $(1,-\frac{1}{2},-2)$, $(0,\frac{1}{2},1)$,  respectively, we expect the ordering $\Delta v_2^p >\Delta v_2^\Lambda > \Delta v_2^{\Xi^-}>\Delta v_2^{K^+}>0$ for reasonable values of $r_S>0$ and $r_I<0$. This tendency is obeyed by most data points except a few in the low energy region. We  note that the $\Xi^-$ data point at $\sqrt{s}=11.5$ GeV should not be taken seriously because, according to the STAR collaboration \cite{Adamczyk:2013gw}, this data point is afflicted with `additional systematic effects which are not included in the error bars'.
 In Fig.~\ref{fig4}, we have also included our prediction for the $\Omega$-baryon. Since $\Omega^-$ has $S=-3$, we expect that $\Delta v_2^{\Omega}\sim \mu_B -3\mu_S$ is smaller  than other baryons.\footnote{Multi-strange hadrons such as $\Omega$ and $\Xi$ may freeze out earlier than non-strange hadrons. Again this uncertainty mostly goes away in the ratio (\ref{from}).}

Concerning the pions, the negative $\Delta v_2^{\pi^+}$ can be naturally explained by the negative isospin chemical potential. However,  the magnitude is problematic. Our choice $r_I=-0.15$, which describes the pion data very well, is too large compared with the value $r_I\approx -0.02\sim -0.03$ extracted from the SPS data
 \cite{BraunMunzinger:1999qy,anton}. We may dial $r_I$ down to, say, $r_I\approx -0.1$ without spoiling much the quality of the $\Delta v_2^{\pi^+}$ fit, but not further down. On the other hand, the other hadrons ($p,\Lambda,\Xi^-,K^+$) are more or less unaffected by $r_I$ and can be well fitted even with $r_I=-0.02$ and $r_S\approx 0.2$.
 This may be an indication that there are other mechanisms to generate the difference $\Delta v_2$ which predominantly act on the pions.\footnote{It is worth mentioning that feed-down corrections (resonance decays) are not included in the current estimates and they could change the fitting parameters.}

  In the large-$\sqrt{s}$ region, our result tends to slightly overestimate $\Delta v_2^X$. This is partly due to  the too fast rise of $v_2^{ideal}$ with energy as mentioned before. However, in Fig.~\ref{fig4} we assumed that $\eta/s=0.2$ is independent of $\sqrt{s}$. A recent hydrodynamic simulation suggests that $\eta/s$ is a decreasing function of $\sqrt{s}$ \cite{Karpenko:2015xea}, and this could alleviate the (small) discrepancy in the large-$\sqrt{s}$ region  (remember that $\Delta v_2^{X}\propto \eta/s$).

\section{Summary and conclusions}

In this paper, we have revealed, in a completely analytical manner, a number of interesting features about the nature of hydrodynamics in the presence of conserved currents as well as the chemical potential (collision energy) dependence of the flow harmonics $v_n$. Let us summarize  the main findings.

\begin{itemize}

\item
As a generalization of the Gubser flow \cite{Gubser:2010ze}, we have derived an anisotropic solution of the relativistic Navier-Stokes equation coupled with conserved currents. Conformal symmetry and boost invariance have been assumed. The solution is valid to linear order in the shear viscosity $\eta$ and the eccentricity $\epsilon_n$. Based on the solution and the Cooper-Frye formula, we analytically computed the flow harmonics $v_n$ at finite density.

\item
In ideal hydrodynamics, the QGP fireball follows a straight line trajectory $\mu/T=const.$ in the phase diagram in the $(T,\mu)$-plane. The shear viscosity causes a deviation from the straight line  as shown in (\ref{alpha}). We expect this picture to be approximately correct in QCD in the deconfined phase of the hydro evolution.


\item
$v_n$ is a decreasing function of density (or an increasing function of $\sqrt{s}$) and decreases faster with $n$ at higher densities $v_n\sim e^{-n\ln(4B^3/27)}$. This is because the lifetime of the hydrodynamic regime ($\sim 1/B^3$) is shorter at high density as it is correlated with the multiplicity $C^3\sim dN/dY$ through the constant energy condition (\ref{kinetic}). In this regard, it is interesting to recall  that in an early numerical study \cite{Kolb:2000sd}, a constant (or even decreasing) $v_2$ as a function of $\sqrt{s}$ was obtained if the hydro simulation is continued to very low temperatures (the so-called `hydro limit' \cite{Voloshin:2002ii}). The rising $v_2$ with energy can be obtained  by switching off hydrodynamics at a relatively high temperature \cite{Teaney:2001av}.  Our assumption of early freezeout is similar in spirit to this.

\item
At finite chemical potential, there are new viscous corrections to $v_n$ (the last terms proportional to $\gamma$ in (\ref{final})). Numerically, they are smaller than the contribution previously found in the $\mu=0$ case \cite{Hatta:2014jva}. However, they give the leading order contribution to the \emph{difference} in $v_n$ between particles and antiparticles.

\item
The viscous corrections to $v_n$ are enhanced at high density. Even if $\eta/s$ is constant, the Knudsen number $K$ grows at high density as it is inversely proportional to the multiplicity (\ref{knud}). At large-$n$, it is also enhanced linearly by $n$, $v_n/v_n^{ideal}\sim 1-{\mathcal O}(nK)$ \cite{Hatta:2014jva}.

\item
The elliptic flow difference between particles and antiparticles $\Delta v_2^X=v_2^X -v_2^{\bar{X}}$ (or more generally, $\Delta v_n^X$) can be nonzero only if the particle $X$ is charged under some chemical potential(s) \emph{and} the shear viscosity is nonvanishing. This is related to the deviation from constancy of the ratio $\mu/T$ due to viscous effects. Our result is summarized by the `master formula' (\ref{master}) (or the more fundamental formula (\ref{from})) which  schematically reads $\Delta v_2^X\propto \eta \mu^X$. This formula dictates the ordering $\Delta v_2^p >\Delta v_2^\Lambda > \Delta v_2^{\Xi^-}>\Delta v_2^{K^+}>0>\Delta v_2^{\pi^+}$ which seems to be borne out by the STAR result except for a few data points.  Our mechanism of generating $\Delta v_2^X$ is distinct from the previous theoretical considerations in \cite{Burnier:2011bf,Xu:2013sta}, but we find it has some common ground with the discussion in \cite{Steinheimer:2012bn}. Finally we pointed out that the observed magnitude of $\Delta v_2^{\pi^+}$ is large  and can be fitted only if we assume an unnaturally large value of the isospin chemical potential $\mu_I$. This suggests that other mechanisms to generate $\Delta v_2^{\pi}$  may be at work.

\end{itemize}
Presumably some of the above features are empirically well known to the experts of hydrodynamic simulations. However, they have not been systematically derived with the level of analytical detail  presented in this work.

There are a number of directions for future work. Admittedly, the assumptions of boost invariance and conformal invariance are too simplistic, especially at high density. One has to relax these approximations to be more realistic. Related to this, we only considered the conformal equation of state $\varepsilon=3p=T^4f(\mu/T)$ where the function $f$ does not carry any information about the crossover and possibly first order phase transitions at finite density. (Nevertheless it is remarkable that we can explain many features of $v_n$ measured at different energies without such information.) It is important to figure out   how the presence of phase transitions in $f$ is encoded in the observed behavior of $v_n$. Including the effects of anomaly (see, e.g., \cite{Hongo:2013cqa}) is also interesting. We hope to address these questions in future work.


\section*{Acknowledgements}
We thank Hiroshi Masui for an incentive remark which partly motivated this work.
We also thank Anton Andronic, Kenji Morita, Guangyou Qin, Fuqiang Wang, Nu Xu and the members of the nuclear theory group of Kyoto University for interesting discussions and helpful comments. A.~M. is supported by the RIKEN Special Postdoctoral Researcher program.

\appendix
\section{Computation of $\delta J_{2,3}$}

In this Appendix, we carry out the computation of $\delta J_{2,3}$ defined in (\ref{j123}).
We first note that the last two terms of (\ref{sur}) may be combined as
\beq
&& -p_T\cos(\phi-\phi_p)\frac{\partial \tau}{\partial x_\perp} + \frac{p_T}{x_\perp}\sin(\phi-\phi_p)\frac{\partial \tau}{\partial \phi}
 \to \frac{(2L)^5p_T}{B^3(L^2+x_\perp^2)^3}\Biggl[ 4x_\perp \cos\phi +3\epsilon_n  \left(\frac{2Lx_\perp}{L^2+x_\perp^2}\right)^n \nn
 && \qquad  \biggl\{ \left(-4x_\perp +n\frac{L^2-x_\perp^2}{x_\perp}\right)  \cos n(\phi+\phi_p) \cos \phi +n\frac{L^2+x_\perp^2}{x_\perp} \sin n(\phi+\phi_p)\sin\phi\biggr\} \Biggr]\,, \label{line}
\eeq
 where we  shifted the integration variable $\phi$ as $\phi\to \phi+\phi_p$.
We then make the following replacement
\beq
\cos n(\phi+\phi_p) \cos \phi \to \cos n\phi \cos \phi \cos n\phi_p &\to& \frac{1}{2}\cos (n-1)\phi \cos n\phi_p\,, \nn
\sin n(\phi+\phi_p)\sin \phi \to \sin n\phi \sin \phi \cos n\phi_p
&\to& \frac{1}{2} \cos (n-1)\phi \cos n\phi_p\,,
\eeq
where we neglected $\sin n\phi_p$ which will vanish after the $\phi$-integral, and also $\cos (n+1)\phi$ which will lead to subleading terms $I_{n+1}(z)\sim z^{n+1}$ after the $\phi$-integral compared to $I_{n-1}(z)\sim z^{n-1}$.
 Thus,   (\ref{line}) effectively becomes
\beq
\frac{(2L)^5p_T}{B^3(L^2+x_\perp^2)^3}\Biggl[ 4x_\perp \cos\phi +3\epsilon_n  \left(\frac{2Lx_\perp}{L^2+x_\perp^2}\right)^n  \left(-2x_\perp +\frac{nL^2}{x_\perp}\right)  \cos (n-1)\phi \cos n\phi_p  \Biggr]\,.
\eeq
Similarly, we can write $\delta U$ defined in (\ref{du}) as
\beq
\delta U&\to & \frac{p_T}{T_c} \left(\delta u_\perp \cos \phi \cos n(\phi+\phi_p) -\frac{\delta u_\phi}{x_\perp}\sin \phi \sin n(\phi+\phi_p)\right) \nn
&\to & \frac{z}{2} \frac{1}{u_{\perp 0}}\left(\delta u_\perp  -\frac{\delta u_\phi}{x_\perp}\right)\cos (n-1)\phi \cos n\phi_p \nn
&=& \frac{3z}{4x_\perp} \left(\frac{2Lx_\perp}{L^2+x_\perp^2}\right)^n \left(-2x_\perp + \frac{nL^2}{x_\perp}\right)\cos (n-1)\phi \cos n\phi_p\,.
\eeq

Using these simplifications, we get
\beq
\delta J_2+\delta J_3&\sim& \epsilon_n e^{k\alpha} \frac{(2L)^6 \gamma K p_T}{B^6} \int d\zeta dx_\perp d\phi \frac{x_\perp}{(L^2+x^2_\perp)^3} e^U \Biggl[ 4x_\perp \delta U\left(\frac{f'}{4f}U+\frac{f'}{4f}+k\right) \cos \phi \nn
&& +3\left(\frac{2Lx_\perp}{L^2+x_\perp^2}\right)^n  \left(U\frac{f'}{4f}+k\right)
\left(-2x_\perp + \frac{nL^2}{x_\perp}\right)\cos (n-1)\phi \cos n\phi_p\Biggr] \nn
&=&\epsilon_n \frac{(2L)^6 \gamma K p_T}{B^6} \int d\zeta dx_\perp d\phi \frac{x_\perp}{(L^2+x^2_\perp)^3} e^U \left(\frac{2Lx_\perp}{L^2+x_\perp^2}\right)^n \left(-2x_\perp + \frac{nL^2}{x_\perp}\right) \nn && \Biggl[ 3z  \left(\frac{f'}{4f}U+\frac{f'}{4f}+k\right) \cos \phi +3  \left(U\frac{f'}{4f}+k\right)\Biggr]\cos (n-1)\phi \cos n\phi_p\,,
\eeq
 where
 \beq
 U=-\frac{m_T}{T_c}\cosh (\zeta-Y) + z\cos \phi\,.
 \eeq
Let us now define
\beq
Y_n(a)&\equiv &\int d\zeta d\phi e^{aU} \cos n\phi = 4\pi K_0(am_T/T_c)I_n(az)\nn
&\approx & 4\pi K_0(am_T/T_c) \frac{1}{n!}\left(\frac{az}{2}\right)^n\,.
\eeq
 Using this we obtain
 \beq
&& \delta J_2+\delta J_3 \approx  \frac{(2L)^6 \gamma K p_T}{B^6} e^{k\alpha} \int  dx_\perp \frac{x_\perp}{(L^2+x^2_\perp)^3}
\left(\frac{2Lx_\perp}{L^2+x_\perp^2}\right)^n \left(-2x_\perp + \frac{nL^2}{x_\perp}\right) \nn && \qquad \qquad \qquad \times \Biggl[ \frac{3z}{2}  \left(\frac{f'}{4f}Y'_{n-2}+\left(\frac{f'}{4f}+k\right)Y_{n-2}\right)  +3  \left(\frac{f'}{4f}Y'_{n-1}+kY_{n-1}\right)\Biggr]_{a=1} \nn
&& \qquad =4\pi \frac{(2L)^6 \gamma K p_T}{B^6} e^{k\alpha} \int  dx_\perp \frac{x_\perp}{(L^2+x^2_\perp)^3}
\left(\frac{2Lx_\perp}{L^2+x_\perp^2}\right)^n \left(-2x_\perp + \frac{nL^2}{x_\perp}\right) \nn && \qquad \qquad \times \frac{3n}{(n-1)!}\left(\frac{z}{2}\right)^{n-1} \left[ \frac{f'}{4f}\left(-\frac{m_T}{T_c}K_1(m_T/T_c) + (n-1)K_0(m_T/T_c)\right) + k K_0(m_T/T_c)\right] \nn
&& \qquad =e^{k\alpha}\frac{36\pi}{B^3}\gamma KT_c\left(\frac{64p_T}{T_cB^3}\right)^n L^3\frac{n^2(n-1)}{3n-1}\frac{\Gamma(3n)}{\Gamma(4n)} \nn
 && \qquad \qquad \times \left[ \frac{f'}{4f}\left(-\frac{m_T}{T_c}K_1(m_T/T_c) + (n-1)K_0(m_T/T_c)\right) + k K_0(m_T/T_c)\right]\,.
\eeq
The correction to $v_n$ can be computed analogously to (\ref{cor}), and the result
 is reported in (\ref{ap}).

\end{document}